\documentclass[12pt,letterpaper]{article}
\usepackage{amsmath}
\usepackage{amssymb}

\def\mod#1#2{
\count255=#1 \divide\count255 by #2
\multiply\count255 by #2
\advance\count255 by -#1
#1=-\count255}
\newcount\hour
\hour=\time \divide\hour by 60
\newcount\minute
\minute=\time \mod{\minute}{60}

\newcommand{\mg}{{\bf g}}
\newcommand{\mf}{{\bf f}}
\newcommand{\mF}{{\bf F}}

\newcommand{\mM}{{\bf M}}
\newcommand{\mN}{{\bf N}}
\newcommand{\mQ}{{\bf Q}}
\newcommand{\mR}{{\bf R}}
\newcommand{\mS}{{\bf S}}
\newcommand{\mZ}{{\bf Z}}
\newcommand{\mRt}{\tilde{\bf R}}
\newcommand{\mQt}{\tilde{\bf Q}}
\newcommand{\mPhi}{{\boldsymbol{\Phi}}}
\newcommand{\mPsi}{{\boldsymbol{\Psi}}}

\newcommand{\tr}{{\mathop{\mathrm{tr}}\nolimits}}
\newcommand{\Tr}{{\mathop{\mathrm{Tr}}\nolimits}}

\newcommand{\bracket}[1]{{\langle{#1}\rangle}}
\newcommand{\Bracket}[1]{{\left\langle{#1}\right\rangle}}

\newcommand{\CF}{{\cal F}}

\newcommand{\CM}{{\cal M}}
\newcommand{\CN}{{\cal N}}
\newcommand{\CO}{{\cal O}}

\newcommand{\CR}{{\cal R}}

\newcommand{\CW}{{\cal W}}

\newcommand{\ft}{{\tilde{f}}}
\newcommand{\gt}{{\tilde{g}}}

\newcommand{\Pt}{{\tilde{P}}}
\newcommand{\Qt}{{\tilde{Q}}}
\newcommand{\Rt}{{\tilde{R}}}
\newcommand{\Tt}{{\tilde{T}}}
\newcommand{\Wt}{{\tilde{W}}}

\newcommand{\lambdat}{{\tilde{\lambda}}}

\newcommand{\Phit}{{\tilde{\Phi}}}

\newcommand{\unit}{{1\hspace{-.8ex}1}}

\newcommand{\so}{\Rightarrow}

\newcommand{\ba}{\begin{eqnarray}}
\newcommand{\ea}{\end{eqnarray}}
\newcommand{\WW}{\mbox{$\cal W$}}

\begin{document}

\begin{flushright}
UCLA/03/TEP/13\\
hep-th/0304138
\end{flushright}

\begin{center}

{\Large\bf 
Loop equations, matrix models, and
\\
$\CN=1$ supersymmetric gauge theories
}\\[5ex]

Per Kraus, Anton V. Ryzhov, and Masaki Shigemori\\[5ex]

{\it Department of Physics and Astronomy,}\\
{\it  UCLA, Los Angeles, CA 90095-1547,}\\
{\tt pkraus, ryzhovav, shige@physics.ucla.edu}\\

\end{center}

\begin{abstract}
We derive the Konishi anomaly equations for $\CN=1$ supersymmetric gauge
theories based on the classical gauge groups with matter in 
two-index tensor and fundamental representations, thus 
extending the existing results for $U(N)$.
A general formula is obtained which expresses  solutions to the Konishi
anomaly equation in terms of solutions to the loop equations of the
corresponding matrix model.
This provides an alternative to the diagrammatic proof that the
perturbative part of the glueball superpotential $W_{\rm eff}$ for these
matter representations can be computed from matrix model integrals, and
further shows that the two approaches always give the same result.  The
anomaly approach is found to be computationally more efficient in the
cases we studied.
%
Also, we show in the anomaly approach how theories with a traceless
two-index tensor can be solved using an associated theory with a
traceful tensor and appropriately chosen coupling constants.

\end{abstract}

\newpage

\section{Introduction}

The recently established connection
\cite{Dijkgraaf:2002fc,Dijkgraaf:2002vw,Dijkgraaf:2002dh} between
matrix models and the effective superpotentials of certain ${\cal
N}=1$ gauge 
 theories provides us with a new tool for studying
 supersymmetric field theories.   The connection, originally
 formulated in the context of $U(N)$ gauge theories with adjoint
 matter, has been established following two distinct approaches, one
 based on superspace diagrammatics \cite{Dijkgraaf:2002xd}, and the other on generalized
 Konishi anomalies \cite{Cachazo:2002ry}.  These derivations were subsequently
 generalized to a few more gauge groups and matter
 representations, but the list of examples is actually quite short at
 present.  In particular, the diagrammatic approach has been
 applied to the classical gauge groups with matter in arbitrary
 two-index representations 
\cite{Ita:2002kx,Ashok:2002bi,Janik:2002nz,Kraus:2003jf,Naculich:2003cz},
 while the anomaly approach has so far
 been used for $U(N)$ with matter in the adjoint and
 fundamental representations 
\cite{Cachazo:2002ry,Seiberg:2002jq}, in (anti)symmetric tensor 
representations \cite{Naculich:2003cz}, and  to quiver theories 
\cite{Naculich:2003cz,Casero:2003gf}.
\footnote{The Konishi anomalies have
also been applied without direct reference to a matrix model in
\cite{Brandhuber:2003va}.}   So basic questions remain
 regarding the general applicability of these ideas, and also
 whether matrix models can in fact successfully reproduce the known
 physics of supersymmetric gauge theories.

In \cite{Kraus:2003jf}, theories based on the classical gauge
groups with two-index tensor matter were considered using the
diagrammatic approach. In the case\footnote{Our
convention is such that $Sp(2) \approx SU(2)$.}  of $Sp(N)$ with
anti-symmetric matter, a comparison was made against an
independently derived dynamical superpotential \cite{Cho:1996bi}
governing these theories.  The comparison revealed agreement up to
$h-1$ loops in perturbation theory ($h$ is the dual Coxeter
number), and a disagreement at $h$ loops and beyond.  Although it
seemed most likely that the disagreement was due to
nonperturbative effects, even at the perturbative level there were
a number of subtleties deserving of further scrutiny. These
subtleties mainly concern the class of diagrams which should be
kept in the evaluation of the superpotential, and whether one is
allowed to use Lie algebra identities to express objects of the
form $\Tr ({\cal W}_\alpha )^{2h}$ in terms of lower traces
including the glueball superfield $S \sim \Tr ({\cal
W}_\alpha)^2$.  Since these subtleties arise at the same order in
perturbation theory as the observed discrepancies, it seems
important to gain a better understanding of them.   One motivation
for the present work was to rederive the results of
\cite{Kraus:2003jf} in the anomaly approach to see if this gives
the same result, and if so, to see which diagrams are effectively
being computed.  We will see that the anomaly approach corresponds
to keeping at most two ${\cal W}_\alpha$'s per index loop and not
using Lie algebra identities. So using these rules, whether one
computes using diagrams or anomalies, one finds the same
agreements/discrepancies between the gauge theory and the matrix
model.

Another motivation for this work was to apply the anomaly approach
to a wider class of theories.  For the classical gauge groups with
certain two-index tensors plus fundamentals, we will show how
solutions to the Konishi anomaly equations can be obtained from
solutions to the loop equations of the corresponding matrix model.  This leads
to the following general formula for the perturbative contribution
to the effective glueball superpotential
\begin{align}
 W_{\rm eff}
 =
 N\frac{\partial}{\partial S}\CF_{S^2}
 +\frac{w_\alpha w^\alpha}{2}\frac{\partial^2}{\partial S^2}\CF_{S^2}
 +4\CF_{RP^2} +\CF_{D^2}
 \label{eqa}
\end{align}
where the ${\cal F}$'s are matrix model contributions of a given
topology to the free energy.   This formula generalizes the $U(N)$
results of \cite{Cachazo:2002ry,Seiberg:2002jq,Naculich:2003cz}, as well as
results 
\cite{Ita:2002kx,Ashok:2002bi,Janik:2002nz,Kraus:2003jf,Naculich:2003cz}
found using the diagrammatic approach.

In fact, the above formula is only directly applicable to cases in
which no tracelessness condition is imposed on the two-index
tensors.    In \cite{Kraus:2003jf} it was shown that imposing a
tracelessness condition requires one to include additional
disconnected matrix model diagrams, and there was no simple
formula relating the superpotential to the free energy of the
traceless matrix model. On the other hand, one expects that the
traceful theory should contain all the information about the
traceless case provided one includes a Lagrange multiplier field
to set the trace to zero. We will show how this works in detail,
and find that indeed, the superpotential of the traceless theory can
be extracted from the free energy of the traceful matrix model. We
use this to rederive and extend some results from 
\cite{Kraus:2003jf} in a much more convenient fashion.

The remainder of this paper is organized as follows.  In sections
2 and 3 we derive the gauge theory Konishi anomaly equations and
the matrix model loop equations for the theories of interest.  The
theories can all be treated in a uniform way by using appropriate
projection operators.  In section 4 we discuss some of the
subtleties alluded to above, and then go on to show that solutions
to the  gauge theory anomaly equations follow from those of the
matrix model loop equations.  Section 5 concerns the effects of
tracelessness.  Details of some of our calculations are given in
appendices A and B.

%

Note: As we were preparing the manuscript, \cite{Alday-Cirafici} 
appeared which overlaps with some of our discussion.

\section{Loop equations on the gauge theory side}

In this section we  derive the gauge theory loop equations for
various gauge groups and matter representations, extending the 
$U(N)$ result of
\cite{Cachazo:2002ry,Seiberg:2002jq}.

\subsection{Setup}

We consider an $\CN=1$ supersymmetric gauge theory with  tree level
superpotential
\begin{align}
 W_{\rm tree}=
  \Tr[W(\Phi)]+\Qt_\ft m_{\ft f}(\Phi) Q_f,
 \label{eq:TL_superpot_def}
\end{align}
where the two-index tensor $\Phi_{ij}$ is in one of the following
representations:
\begin{itemize}
 \item $U(N)$ adjoint.
 \item $SU(N)$ adjoint.
 \item $SO(N)$ antisymmetric tensor.
 \item $SO(N)$ symmetric tensor, traceful or traceless.
 \item $Sp(N)$ symmetric tensor.
 \item $Sp(N)$ antisymmetric tensor, traceful or traceless.
\end{itemize}
For other $U(N)$ representations, see \cite{Klemm:2003cy,Naculich:2003cz}.
In the $Sp$ cases, the object with the denoted symmetry is related to $\Phi$ by
\begin{align}
 \Phi=
 \begin{cases}
  SJ & \text{$S_{ij}$: symmetric tensor},\\
  AJ & \text{$A_{ij}$: antisymmetric tensor}.
 \end{cases}
\end{align}
Here $J$ is the invariant antisymmetric tensor of $Sp(N)$, namely
\begin{align}
 J_{ij}=\begin{pmatrix}0&\unit_{N/2}\\-\unit_{N/2}&0\end{pmatrix}.
\end{align}
The tracelessness of the $Sp$ antisymmetric tensor is defined with
respect to this $J$, i.e., by $\Tr[AJ]=0$.

Also, $Q_f$ and $\Qt_f$ are fundamental matter fields, with $f$ and
$\ft$ being flavor indices.  In the $U(N)$ case we have $N_f$
fundamentals $Q_f$ and $N_f$ anti-fundamentals $\Qt_{\ft}$, while in the
$SO/Sp$ case we have 
$N_f$ fundamentals $Q_f$.  In the $SO/Sp$ case, $\Qt_\ft$ is
not an independent field but related to $Q_f$ by
\begin{align}
 (\Qt_{\ft})_i
 =
 \begin{cases}
  (Q_{\ft})_i & SO(N),\\
  (Q_{\ft})_j J_{ji} & Sp(N).
 \end{cases}
 \label{eq:def_Qtilde}
\end{align}
In the $Sp$ case, $N_f$ should be taken to be even to avoid the Witten
anomaly \cite{Witten:fp}.

$W$ and $m$ are taken to be polynomials
\begin{align}
 W(z)=\sum_{p=1}^n\frac{g_p}{p}z^p, \qquad
 m_{\ft f}(z)=\sum_{p=1}^{n'} \frac{(m_p)_{\ft f}}{p}z^p,
\end{align}
where in the traceless cases the $p=1$ term is absent from $W(z)$.  
Further, due to
the symmetry properties of the matrix $\Phi$, some  $g_p$ vanish for
certain representations:
\begin{align}
 g_{2p+1}=0 \quad (p=0,1,2,\cdots)
 \qquad\text{for $SO$ antisymmetric / $Sp$ symmetric}.
\end{align}
The symmetry properties of $\Phi$ also imply that the matrices
$(m_p)_{\ft f}$ have the following symmetry properties:
\begin{align}
 (m_p)_{f'f}
 =
\begin{cases}
  (-1)^p(m_p)_{ff'}     & \text{$SO$ antisymmetric,} \\
  (m_p)_{ff'}           & \text{$SO$ symmetric,} \\
  (-1)^{p+1}(m_p)_{ff'} & \text{$Sp$ symmetric,} \\
  -(m_p)_{ff'}          & \text{$Sp$ antisymmetric.}
\end{cases}
 \label{eq:sym_prop_m}
\end{align}

In this and the next few sections, we discuss traceful cases only,
postponing the traceless cases to section \ref{sec:traceless_cases} 
(we regard the $SU(N)$ case as the traceless $U(N)$ case).

\subsection{The loop equations}
\label{section: loop equations}

%

We will be interested in expectation values of chiral operators.  
As in \cite{Cachazo:2002ry,Seiberg:2002jq}, 
\begin{align}
 \{\CW_\alpha,\CW_\beta\}
=
 [\Phi,\CW_\alpha]
=
 \CW_\alpha Q
=
 \Qt \CW_\alpha
=
 0
 \label{eq:chiral_ring_relations}
\end{align}
in the chiral ring. 
Therefore, the complete list of independent single-trace chiral
operators are $\Tr[\Phi^p]$, $\Tr[\CW_\alpha\Phi^p]$,
$\Tr[\CW^2\Phi^p]$, and $\Qt_\ft \Phi^p Q_f$. 
As is standard, we define
\begin{align}
 S= -\frac{1}{32\pi}\Tr[\CW_\alpha \CW^\alpha],
\quad
w_\alpha=\frac{1}{4\pi}\Tr[\CW_\alpha].
\end{align}
The chiral operators can be packaged
concisely in terms of the resolvents
\begin{gather}
 R(z)
 \equiv
 -\frac{1}{32\pi^2} \Bracket{\Tr\left[\frac{\CW^2}{z-\Phi}\right]},
 \quad
 w_\alpha(z)  \equiv  \frac{1}{4\pi}
 \Bracket{\Tr\left[\frac{\CW_\alpha}{z-\Phi}\right]},
 \quad
 T(z)
 \equiv
 \Bracket{\Tr\left[\frac{1}{z-\Phi}\right]},
 \nonumber\\
 M_{f\ft}(z)
 \equiv
 \Bracket{\Qt_\ft \frac{1}{z-\Phi} Q_f}.
 \label{eq:def_resolvent_gt}
\end{gather}
Note that the indices of $M_{f\ft}$ are reversed relative to
$\Qt_\ft$, $Q_f$.  The resolvent $w_\alpha(z)$ is nonvanishing only
for $U(N)$; in all other cases $w_\alpha(z)\equiv 0$. This can be
understood as follows.  In these semi-simple cases the Lie algebra
generators are traceless, so we cannot have a nonzero background field
$w_\alpha$.  There being no preferred
spinor direction specified by the background $w_\alpha$, the spinor
$w_\alpha(z)$ can be nothing but zero.  Alternatively, if we integrate
out $\Phi$, then $w_\alpha(z)$ should be of the form 
$\Bracket{\Tr[\CW_\alpha](\Tr[\CW^2])^n}$ by  the chiral ring relations
(\ref{eq:chiral_ring_relations}).  If we use the factorization property
of chiral operator expectation values, this is proportional to $w_\alpha
S^n$, which vanishes.

The resolvents defined in equation (\ref{eq:def_resolvent_gt}) 
provide sufficient data to determine the effective
superpotential up to a coupling independent part, because of the
relation
\begin{align}
 \Bracket{\Tr[\Phi^p]}
 = p\frac{\partial}{\partial g_p}W_{\rm eff},
 \qquad
 \Bracket{\Qt_{\ft} \Phi^p Q_f}
 =
 p\frac{\partial}{\partial (m_p)_{\ft f}}W_{\rm eff}.
 \label{eq:rel_Weff_and_trPhi_etc}
\end{align}

The generalized Konishi \cite{Konishi:1983hf} anomaly equation
\cite{Cachazo:2002ry,Seiberg:2002jq,Brandhuber:2003va} is obtained by
considering the divergence of the current associated with the variation
of a particular field $\Psi_a$:
\begin{align}
 \delta \Psi_a=f_a,
 \label{eq:gK_anomaly_variation}
\end{align}
where $a$ is a gauge index.  Then the anomaly equation reads
\begin{align}
 \Bracket{\frac{\partial W_{\rm tree}}{\partial \Psi_a} f_a}
 +
 \frac{1}{32\pi^2}
 \Bracket{[\CW_\alpha \CW^\alpha]_a^b \frac{\partial f_b}{\partial \Psi_a}}
 =0,
 \label{eq:gK_anomaly}
\end{align}
where $\CW_\alpha$ is in the representation furnished by  $\Psi$.
The first term in (\ref{eq:gK_anomaly}) 
represents the classical change of the action under the variation
(\ref{eq:gK_anomaly_variation}), 
while the second term in (\ref{eq:gK_anomaly}) corresponds to
the quantum variation due to the change in the functional measure.

In the $U(N)$ case considered in \cite{Cachazo:2002ry,Seiberg:2002jq}, 
there is no additional symmetry imposed on the field $\Phi$, 
so $\delta \Phi_{ij}=f_{ij}$ can be any function of
$\CW_\alpha$ and $\Phi$. 
In general, the tensor $\Phi$ will have some symmetry properties 
(symmetric or antisymmetric tensor in the present $SO/Sp$ study),
and $f_{ij}$ should
be chosen to reflect those. 
Similarly, the derivative 
$\partial/\partial \Psi_a=\partial/\partial \Phi_{ij}$ 
should be defined in accord with the symmetry property of $\Phi_{ij}$.
To this end, 
we define a projector $P$ 
appropriate to
each case: 
\begin{align}
 P_{ij,kl}=
  \frac{1}{2}(\delta_{ik}\delta_{jl}+\sigma t_{il} t_{jk}) ,
 \label{eq:def_projector}
\end{align}
where
\begin{align}
 \begin{cases}
  t_{ij}=\delta_{ij},~ \sigma=-1 & \text{$SO$ antisymmetric,}\\
  t_{ij}=\delta_{ij},~ \sigma=+1 & \text{$SO$ symmetric,    }\\
  t_{ij}=J_{ij}     ,~ \sigma=-1 & \text{$Sp$ symmetric,    }\\
  t_{ij}=J_{ij}     ,~ \sigma=+1 & \text{$Sp$ antisymmetric }.
 \end{cases}
 \label{eq:def_sigma_and_t}
\end{align}
The tensor $\Phi_{ij}$ satisfies $P_{ij,kl}\Phi_{kl}=\Phi_{ij}$.  Then,
the symmetry property of $\delta \Phi$ discussed above is implemented by
the replacements
\begin{align}
 f_{a} = f_{ij}\to  P_{ij,kl}f_{kl},
 \qquad
 \frac{\partial}{\partial \Psi_{a}}=
 \frac{\partial}{\partial \Phi_{ij}}\to
 P_{ij,kl}\frac{\partial}{\partial \Phi_{kl}}.
 \label{eq:act_with_projectors_gt}
\end{align}
With this replacement, $f_{ij}$ can be any function of $\CW_\alpha$ and
$\Phi$ as in the $U(N)$ case.  The derivative can be treated as in
the $U(N)$ case also.

There is no such  issue for the $Q$ and $\Qt$ fields, although we have to
remember that they are not independent for $SO/Sp$.

With the projectors in hand, there is no difficulty in
deriving the loop equations for $SO/Sp$.  Here we just present the
resulting loop equations, leaving the details to Appendix
\ref{app:loop_eq_gt}:
\begin{equation}
\begin{split}
 [W'R]_-&= \tfrac{1}{2} R^2, \\
 [W'T+\tr(m'M)]_-&=
 \begin{cases}
  \left(T-\frac{2}{z}\right)R    & \text{$SO$ antisymmetric}, \\
  \left(T-2\frac{d}{dz}\right)R  & \text{$SO$ symmetric    }, \\
  \left(T+\frac{2}{z}\right)R    & \text{$Sp$ symmetric    }, \\
  \left(T+2\frac{d}{dz}\right)R  & \text{$Sp$ antisymmetric},
 \end{cases}
 \\
 2[(Mm)_{ff'}]_-     &=  R\delta_{ff'    }, \\
 2[(mM)_{\ft\ft'}]_- &=  R\delta_{\ft\ft'},
\end{split}
\label{eq:loop_eq_gt}
\end{equation}
where $[F(z)]_-$ means to drop non-negative powers in a Laurent
expansion in $z$.  The last two equations are really the same equation
due to the symmetry properties of 
$m$ (see equation (\ref{eq:sym_prop_m})), and $\Phi$.  
Note that there is no $w_\alpha(z)$ in these
cases as explained below Eq.\ (\ref{eq:def_resolvent_gt}).
For the sake of comparison, the $U(N)$ loop equations  are
\cite{Cachazo:2002ry,Seiberg:2002jq}
\begin{equation}
\begin{split}
 [W'R]_-&=   R^2,\\
 [W'w_\alpha]_-&=2w_\alpha R,\\
 [W'T+\tr(m'M)]_-&=  2TR+w_\alpha w^\alpha\\
 [(Mm)_{ff'}]_-    &= R\delta_{ff'},\\
 [(mM)_{\ft\ft'}]_- &= R\delta_{\ft\ft'}.
\end{split}
\label{eq:loop_eq_gt_U(N)}
\end{equation}
One observes some extra numerical factors in the $SO/Sp$ case as
compared to the $U(N)$ case.  The $\frac{1}{2}$ in the first equation is
from the $\frac{1}{2}$ in the definition of $P_{ij,kl}$, while the
factor $2$ in the last two equations is because in the $SO/Sp$ case $Q$
and $\Qt$ are really the same field, so the variation of $\Qt mQ$ under
$\delta Q$ for $SO/Sp$ is twice as large as that for $U(N)$.  Finally,
the $\frac{1}{z}R(z)$ and $\frac{d}{dz} R(z)$ 
terms in the second equation of
(\ref{eq:loop_eq_gt}) come from the second term of $P_{ij,kl}$.

The solution to the loop equations (\ref{eq:loop_eq_gt}) or
(\ref{eq:loop_eq_gt_U(N)}) is determined uniquely \cite{Cachazo:2002ry}
given the condition
\begin{align}
 S
 =\oint_{C} \frac{dz}{2\pi i} R(z),\qquad
 w_\alpha
 =\oint_{C} \frac{dz}{2\pi i} w_\alpha(z),\qquad
 N
 =\oint_{C} \frac{dz}{2\pi i} T(z),
 \label{eq:requirement_for_gt_resolvents}
\end{align}
where the second equation is only for the $U(N)$ case.  The contour $C$
goes around the critical point of $W(z)$.  Therefore, if we
recall the relation (\ref{eq:rel_Weff_and_trPhi_etc}), we can say that
the loop equations are all we need to determine the superpotential
$W_{\rm eff}$.

\section{Loop equations on the matrix model side}
\label{section: mm loop equations}

Let us consider the matrix model which corresponds to the gauge theory
in the previous section.  
Its partition function is 
\begin{align}
 \mZ
 =
 e^{-\frac{1}{\mg^2}\mF(\mS)}
 =
 \int d\mPhi d\mQ d\mQt \,
   e^{-\frac{1}{\mg}W_{\rm tree}(\mPhi,\mQ,\mQt)}.
 \label{eq:mm_def}
\end{align}
We denote matrix model quantities by boldface letters.  Here, $\mPhi$ is an
$\mN\times\mN$ matrix with the same symmetry property as the
corresponding matter field in the gauge theory.  $\mQ_f$ and $\mQt_{\ft}$
are defined in a similar way to their gauge theory counterparts
(therefore $d\mQt$ in (\ref{eq:mm_def}) is not included for $SO/Sp$).
The function (or the ``action'') $W_{\rm tree}$ is the one defined in
(\ref{eq:TL_superpot_def}).  We will take the $\mN\to\infty$, $\mg\to 0$
limit with the 't Hooft coupling $\mS=\mg \mN$ kept fixed.  The dependence
of the free energy $\mF(\mS)$ on $\mN$ is eliminated using the relation
$\mN=\mS/\mg$, and we expand $\mF(\mS)$ as
\begin{align}
 \mF(\mS)
 =\sum_{\CM} \mg^{2-\chi(\CM)} \CF_{\CM}(\mS)
 =\CF_{S^2}+\mg\CF_{RP^2}+\mg\CF_{D^2}+\cdots,
\end{align}
where the sum is over all compact topologies $\CM$ of the matrix model
diagrams written in the 't Hooft double-line notation, and $\chi(\CM)$ is
the Euler number of $\CM$.  The cases which will be of interest to us are
the sphere $S^2$, projective plane $RP^2$, and  disk $D^2$, with
$\chi=2,1$, and 1, respectively.  All other contributions have
$\chi\le 0$.

We define matrix model resolvents as follows:
\begin{align}
 \mR(z)
 \equiv
 \mg\Bracket{\Tr\left[\frac{1}{z-\mPhi}\right]}
 ,\qquad
 \mM_{f\ft}(z)
 \equiv
 \mg\Bracket{\mQt_\ft \frac{1}{z-\mPhi} \mQ_f}.
 \label{eq:def_mm_resolvents}
\end{align}
These resolvents provide sufficient data to determine the free energy $\mF$
up to a coupling independent part since
\begin{align}
 \mg\Bracket{\Tr[\mPhi^p]}
 = p \frac{\partial}{\partial g_p}\mF,\qquad
 \mg\Bracket{\mQt_{\ft}\mPhi^p\mQ_f}
 =  p \frac{\partial}{\partial (m_p)_{\ft f}}\mF.
\end{align}
We expand the resolvents in topologies just as we did for $\mF$:
\begin{align}
 \mR(z)=\sum_{\CM} \mg^{2-\chi(\CM)} \mR_\CM(z),
 \qquad
 %
 \mM(z)=\sum_{\CM} \mg^{2-\chi(\CM)} \mM_\CM(z).
 \label{eq:mm_resolvents_expansion}
\end{align}
Although $\mR_{RP^2}$ and $\mR_{D^2}$ are of the same order in $\mg$,
they can be distinguished unambiguously because all terms in $\mR_{D^2}$
contains coupling constants $m_{\ft f}$, while $\mR_{RP^2}$ does not
depend on them at all.  This is easily seen in the  diagrammatic expansion of
$\mF$.  Also, because $\mF_{S^2}$ and $\mF_{RP^2}$ do not contain $m$,
the expansion of $\mM$ starts from the disk contribution, $\mM_{D^2}$.

Now we can derive the matrix model loop equations.  Consider changing the
integration variables as
\begin{align}
 \delta \mPsi_a=\mf_a.
 \label{eq:mm_variation}
\end{align}
Since the partition function is invariant under this variation, we
obtain
\begin{align}
 0=
 -\frac{1}{\mg}
 \frac{\partial W_{\rm tree}}{\partial \mPsi_a} \mf_a
 +\frac{\partial \mf_a}{\partial \mPsi_a}.
 \label{eq:mm_anomaly_eq}
\end{align}
The first term came from the change in the ``action'' and corresponds to
the first term (the classical variation) of the generalized Konishi
anomaly equation (\ref{eq:gK_anomaly}).  On the other hand, the second
term came from the Jacobian and corresponds to the second term (the
anomalous variation) of Eq.\ (\ref{eq:gK_anomaly}).

The derivation of the loop equations now can be done exactly in parallel
to the derivation of the gauge theory loop equations.  In the $SO/Sp$
case, we again have to consider the projector $P_{ij,kl}$.  Here we
leave details of the derivation to Appendix \ref{app:loop_eq_mm} and present
the results.  For $SO/Sp$, they are
\begin{gather}
 \mg \Bracket{ \Tr {W'(\mPhi) \over z - \mPhi} }
 + \mg \Bracket{\mQt {m'(\mPhi) \over z - \mPhi} \mQ }
 =
 {1\over 2}
 \Bracket{ \left(\mg\Tr {1 \over z - \mPhi}\right)^{\!\!2} }
 \pm
 {\sigma\over 2} \mg^2
 \Bracket{ \Tr {1 \over (z - \mPhi)(z - \sigma \mPhi)}},
 \nonumber\\
 2\Bracket{ \mQt_{\ft} {m_{\ft f}(\mPhi)  \over z - \mPhi} \mQ_{f'} }
 =
 \mg
 \Bracket{  \Tr {1 \over z - \mPhi} }
 \delta_{ff'},
 \quad
 2\Bracket{ \mQt_{\ft} {m_{\ft' f}(\mPhi)  \over z - \mPhi} \mQ_{f} }
 =
 \mg
 \Bracket{  \Tr {1 \over z - \mPhi} }
 \delta_{\ft\ft'},
 \label{eq:exact_mm_loop_eqs}
\end{gather}
in the $SO$ and $Sp$ cases, respectively.
The last two
equations are really the same because of the symmetry properties of
$\mPhi$ and $m_{\ft f}$.

Equations (\ref{eq:exact_mm_loop_eqs}) 
include terms of all orders in $\mg$.  
Expanding the matrix model expectation values in powers 
of $\mg$, 
plugging in the expansion
(\ref{eq:mm_resolvents_expansion}) and comparing the $\CO(1)$ and
$\CO(\mg^1)$ terms, we obtain the $SO/Sp$ loop equations%
\footnote{In
the $SO$ antisymmetric and $Sp$ symmetric cases, $\mR_{RP^2}$ can be
expressed \cite{Janik:2002nz,Ashok:2002bi} in terms of $\mR_{S^2}$,
which leads to the expression
\begin{align}
 \CF_{S^2}(\mS)=\mp \frac{1}{2} \frac{\partial}{\partial \mS}\CF_{RP^2}
\end{align}
in the $SO$ and $Sp$ cases, respectively.
}.
This is done in Appendix \ref{app:loop_eq_mm}, 
and the results are: 
\begin{equation}
 \begin{split}
 [W'\mR_{S^2}]_- &= \tfrac{1}{2}(\mR_{S^2}{})^2
 \\
 [W'\mR_{RP^2}]_-
 &=
 \begin{cases}
  \left(\mR_{RP^2}-\frac{1}{2z}\right)\mR_{S^2}           & \text{$SO$ antisymmetric} \\
  \left(\mR_{RP^2}-\frac{1}{2}\frac{d}{dz}\right)\mR_{S^2}& \text{$SO$ symmetric} \\
  \left(\mR_{RP^2}+\frac{1}{2z}\right)\mR_{S^2}           & \text{$Sp$ symmetric} \\
  \left(\mR_{RP^2}+\frac{1}{2}\frac{d}{dz}\right)\mR_{S^2}& \text{$Sp$ antisymmetric}
 \end{cases}
  \\
  [W'\mR_{D^2}+\tr(m'\mM_{D^2})]_-  &= \mR_{D^2} \mR_{S^2}
  \\
  2[(\mM_{D^2} m)_{ff'}]_-     &= \mR_{S^2} \delta_{ff'}
  \\
  2[(m \mM_{D^2})_{\ft\ft'}]_- &= \mR_{S^2} \delta_{\ft \ft'} ,
\end{split}
\label{eq:loop_eq_mm}
\end{equation}
We separated the $\mR_{RP^2}$ and $\mR_{D^2}$ contributions using the
difference in their dependence on $m_{\ft f}$ (see the argument below
Eq.\ (\ref{eq:mm_resolvents_expansion})).  
Again, the last two equations are really the same equation.  
For comparison, the $U(N)$ loop
equations are
\begin{equation}
 \begin{split}
 [W'\mR_{S^2}]_- &= (\mR_{S^2}{})^2
 \\
 [W'\mR_{D^2}+\tr(m'\mM_{D^2})]_-  &= 2\mR_{D^2} \mR_{S^2}
 \\
 [(\mM_{D^2} m)_{ff'}]_-     &= \mR_{S^2} \delta_{ff'}
  \\
 [(m \mM_{D^2})_{\ft\ft'}]_- &= \mR_{S^2} \delta_{\ft \ft'} ,
\end{split}
\label{eq:loop_eq_mm_U(N)}
\end{equation}
Note that there is no $RP^2$ contribution for $U(N)$.

The solutions to  equations (\ref{eq:loop_eq_mm}) or
(\ref{eq:loop_eq_mm_U(N)}) are determined uniquely given the condition
\begin{align}
 \mS=\oint_{C} \frac{dz}{2\pi i} \mR_{S^2}(z),\qquad
 0=\oint_{C} \frac{dz}{2\pi i} \mR_{RP^2}(z),\qquad
 0=\oint_{C} \frac{dz}{2\pi i} \mR_{D^2}(z).
 \label{eq:requirement_for_mm_resolvents}
\end{align}
In this sense, the loop equations are all we need to determine the free
energy $\mF$.

\section{Connection between gauge theory and matrix model resolvents}

On the gauge theory side we have arrived at the loop equations
(\ref{eq:loop_eq_gt}).  If we can solve these equations for the
resolvents, in particular for $T(z)$, we will have sufficient data to 
determine the glueball superpotential $W_{\rm eff}(S)$ up to a coupling
independent part. In \cite{Cachazo:2002ry}, it was shown for $U(N)$
with adjoint matter that the solution can be obtained with the help of
an auxiliary matrix model. On the other hand, in
\cite{Dijkgraaf:2002xd,Ita:2002kx,Kraus:2003jf,Naculich:2003cz}
it was proved by perturbative diagram expansion that, for $U(N)$ and
$SO/Sp$ with two-index tensor matter, if one only inserts up to two field
strength superfield $\CW_\alpha$'s per     index loop then the
calculation of $W_{\rm eff}(S)$ reduces to matrix integrals.


However, there are a number of reasons to study further the relation
between the gauge theory and matrix model loop equations.
First, as pointed out in \cite{Kraus:2003jf} (see also p.11 of
\cite{Cachazo:2002ry}, and \cite{Witten:2003ye}), there are subtleties
in using chiral ring relations at order $S^{h}$ and higher, where $h$ is
the dual Coxeter number of the gauge group, and these could be related
to the discrepancies observed in \cite{Kraus:2003jf}.  Since traces of
schematic form $\Tr[(\CW_\alpha^2)^n]$ $(n\ge h)$ can be rewritten in
terms of lower power traces at these orders, imposing chiral ring
relations {\em before\/} using the equation of motion of $S$ is not
necessarily justified.  So, it is important to clarify how this subtlety
is treated in the Konishi anomaly approach.
%
Second, as a practical matter, the anomaly approach is more efficient
than the diagrammatic approach in the cases we studied.


So, let us adopt the following point of view
(some related  ideas were explored in \cite{Brandhuber:2003va}).
Let us not assume the reduction to a matrix model {\it a priori}.  Then 
the gauge
theory resolvents $R$, $T$, and $M$ are just unknown functions that
enable us to determine the coupling dependent part of the glueball
effective action.  We do know that we can evaluate the perturbative
contribution to them by Feynman diagrams, but we do not know
whether they are affected by nonperturbative effects or whether they can
be calculated using a matrix model.  These resolvents satisfy the loop
equations (\ref{eq:loop_eq_gt}), and given the conditions
(\ref{eq:requirement_for_gt_resolvents}), they are determined uniquely.
Similarly, the matrix model resolvents $\mR_{S^2}$,
$\mR_{RP^2}$, $\mR_{D^2}$, and $\mM_{D^2}$ are now just functions
satisfying matrix model loop equations (\ref{eq:loop_eq_mm}).  If we
impose the condition (\ref{eq:requirement_for_mm_resolvents}), these
resolvents are also determined uniquely, and by definition can be 
evaluated in matrix model perturbation theory.

Now, let us ask what the relation between the two sets of resolvents is.
Actually it is simple: if we know the matrix model resolvents, we can
construct the gauge theory resolvents as follows.  In the $SO/Sp$ case,
\begin{equation}
 \begin{split}
 R(z)
 &=
 \mR_{S^2}(z),
 \\
 T(z)
 &=
 N\frac{\partial}{\partial \mS}\mR_{S^2}(z)
 +4\mR_{RP^2}(z) +\mR_{D^2}(z),
 \\
 M(z)&=\mM_{D^2}(z)
 \end{split}
 \label{eq:relation_b/t_resolvents}
\end{equation}
with $S$ and $\mS$ identified; 
in the $U(N)$ case, we get 
\begin{equation}
 \begin{split}
 R(z) &= \mR_{S^2}(z),
 \qquad
 w_\alpha(z) = w_\alpha\frac{\partial}{\partial \mS}\mR_{S^2}(z),
 \\
 T(z)
 &=
 N\frac{\partial}{\partial \mS}\mR_{S^2}(z)
 +\frac{w_\alpha w^\alpha}{2}\frac{\partial^2}{\partial \mS^2}\mR_{S^2}(z)
 +\mR_{D^2}(z),
 \\
 M(z)&=\mM_{D^2}(z)
 \end{split}
 \label{eq:relation_b/t_resolvents_U(N)}
\end{equation}
with the same $S = \mS$ identification.\footnote{Some of these relations
have been written down in 
\cite{Gopakumar:2002wx,Naculich:2002hi,Naculich:2002hr}.}
One can easily check that if the matrix model resolvents satisfy the
matrix model loop equations (\ref{eq:loop_eq_mm}) or
(\ref{eq:loop_eq_mm_U(N)}), then the gauge theory resolvents satisfy the
gauge theory loop equations (\ref{eq:loop_eq_gt}) or
(\ref{eq:loop_eq_gt_U(N)}).  The requirement
(\ref{eq:requirement_for_gt_resolvents}) is also satisfied provided that
the matrix model resolvents satisfy the requirement
(\ref{eq:requirement_for_mm_resolvents}).  Further, these relations lead
to
\begin{align}
 \bracket{\Tr[\Phi^p]}_\text{gauge theory}
 =p\frac{\partial}{\partial g_p}W_{\rm eff}
 =p\frac{\partial}{\partial g_p}
 \left[
 N\frac{\partial}{\partial S}\CF_{S^2}
 +\frac{w_\alpha w^\alpha}{2}\frac{\partial^2}{\partial S^2}\CF_{S^2}
 +4\CF_{RP^2} +\CF_{D^2}
 \right],
\end{align}
which implies a relation between the effective superpotential and the
matrix model quantities:
\begin{align}
 W_{\rm eff}
 =
 N\frac{\partial}{\partial S}\CF_{S^2}
 +\frac{w_\alpha w^\alpha}{2}\frac{\partial^2}{\partial S^2}\CF_{S^2}
 +4\CF_{RP^2} +\CF_{D^2}
 \label{eq:Weff_ito_mm_F}
\end{align}
up to a coupling independent additive part.  This proves that the gauge
theory diagrams considered in the Konishi anomaly approach reduce to
matrix model integrals for all matter representations considered.
Further, we do not have to take into account nonperturbative effects,
since we can assume a perturbative expansion in the matrix model (although,
strictly speaking, one should also  verify that the Konishi anomalies
receive no nonperturbative corrections \cite{Brandhuber:2003va}).

The relations (\ref{eq:relation_b/t_resolvents}) and 
(\ref{eq:relation_b/t_resolvents_U(N)})  are consistent with inserting
at most two ${\cal W}_\alpha$'s per index loop, but not with 
inserting more than two and then using Lie algebra relations.  For instance,
this can be seen from the diagrammatic expansion of $\mR_{S^2}(z)$.  So
this shows us explicitly which diagrams are being computed in the
Konishi anomaly approach.  

In the $U(N)$ case \cite{Cachazo:2002ry}, it was convenient to collect all
the gauge theory resolvents into a ``superfield'' $\CR$, because of the
``supersymmetry'' under a shift of $\CW_\alpha$ by a Grassmann number, and
one could relate $\CR$ to the matrix model resolvent $\mR_{S^2}$.  This
fact enabled one to extract all the gauge theory resolvents solely from
$\mR_{S^2}$. However, in more general cases this trick does not work,
and we have to relate the two sets of resolvents directly as in
(\ref{eq:relation_b/t_resolvents}).

\section{Traceless cases}
\label{sec:traceless_cases}

So far, we considered two-index traceful matter $\Phi_{ij}$, and
discussed the relation between the gauge theory and the
corresponding matrix model.  In this section, we consider
traceless\footnote{In this section, we denote traceless quantities
by tildes to distinguish them from their traceful counterparts.}
tensors $\Phit_{ij}$.
These traceless tensors were studied in \cite{Kraus:2003jf}, and a
method of evaluating the glueball effective superpotential
$\Wt_{\rm eff}(S)$ from the combinatorics of the matrix model
diagrams was given. However, the precise connection between the
gauge theory and the matrix model quantities was not transparent,
since one had to keep some of the matrix model diagrams and drop
others in a way that seemed rather arbitrary from the matrix model
point of view.  Instead, here we show that the calculation of
$\Wt_{\rm eff}(S)$ in gauge theory with {\em traceless\/} matter
reduces to a {\em traceful\/} matrix model.

\subsection{Traceless gauge theory vs.\ traceful matrix model}

To derive the generalized Konishi anomaly equation for a traceless
tensor we have to use the appropriate projector
\begin{align}
 \Pt_{ij,kl}
 \equiv
 P_{ij,kl}-\frac{1}{N}\delta_{ij}P_{mm,kl}
 =P_{ij,kl}-\frac{1}{N}\delta_{ij}\delta_{kl},
\end{align}
where $P$ is the projector of the corresponding traceful theory; 
the second equality holds for any projector defined in 
(\ref{eq:def_projector}).
The anomaly term (the second term of Eq.\ (\ref{eq:gK_anomaly}))
is the same as in the traceful case, since the trace
part is a singlet and does not couple to the gauge field.  Therefore, the
only difference in the anomaly equation between traceful and traceless
cases is in the
classical variation (the first term of Eq.\ (\ref{eq:gK_anomaly})), namely
\begin{align}
 \Tr[(Pf) \Wt'(\Phi)]
 \to
 \Tr[(\Pt f) \Wt'(\Phi)]
 =
 \Tr[(P f) \Wt'(\Phi)] - \frac{1}{N}\Tr[f]\Tr[\Wt'(\Phi)].
 \label{eq:diff_b/t_traceful_and_traceless}
\end{align}

For definiteness, let us focus on $SU(N)$ adjoint matter, which can be
thought of as traceless $U(N)$ adjoint matter, without fundamentals
added;  
we will generalize the discussion to other groups and 
matter representations afterward. 
In this case, the last term of Eq.\
(\ref{eq:diff_b/t_traceful_and_traceless}) changes the $U(N)$ loop
equation (the first and the third lines of (\ref{eq:loop_eq_gt_U(N)}))
to
\begin{align}
 [\Wt'(z)\Rt(z)]_-+g_1 \Rt(z)= \Rt(z)^2,
 \qquad
 [\Wt'(z)\Tt(z)]_-+g_1 \Tt(z)= 2\Rt(z)\Tt(z).
\end{align}
Note that $w_\alpha(z)=0$ for $SU(N)$.  The constant $g_1$ is
\begin{align}
 g_1\equiv -\frac{1}{N}\Bigl\langle\Tr[\Wt'(\Phit)]\Bigr\rangle.
 \label{eq:def_g1t}
\end{align}
If we define
\begin{align}
 W(z)\equiv \Wt(z)+g_1 z,
 \label{eq:shifted_supoerpot_traceless}
\end{align}
the above equations are
\begin{align}
 [W'(z)\Rt(z)]_- = \Rt(z)^2,\qquad
 [W'(z)\Tt(z)]_- = 2\Rt(z)\Tt(z).
 \label{eq:traceless_loop_eq_gt}
\end{align}
These are of the same form as the loop equations with
{\em traceful\/} matter and the tree level superpotential $W$.  
Therefore, in order to obtain the
effective glueball superpotential $\Wt_{\rm eff}(S)$ for traceless
matter, we can instead solve the traceful theory with the shifted tree
level superpotential $W$, choosing the value of $g_1$ appropriately.
The solution to these loop equations is determined uniquely given the
condition
\begin{align}
 S=\oint_{C}\frac{dz}{2\pi i}\Rt(z),\qquad
 N=\oint_{C}\frac{dz}{2\pi i}\Tt(z).
 \label{eq:contour_cond}
\end{align}
In the case of traceful matter, the contour is around a critical
point of the tree level superpotential.  However, for traceless
matter, the loop equations above tell us that the contour should
be taken around the critical point of the shifted superpotential
(\ref{eq:shifted_supoerpot_traceless}), rather than the original
$\Wt$. This is because we cannot change all the eigenvalues of
$\Phit$ independently due to the tracelessness condition
$\Tr[\Phit]=0$.

Let the resolvents of the traceful theory with tree level
superpotential $W(\Phi)$ be $R$ and $T$, with $g_1$ treated as an
independent variable.  $R$ and $T$ are functions of $z$, $g_{p\ge
1}$ as well as $S$, $N$: $R=R(z;g_{p\ge 1},S)$, $T=T(z;g_{p\ge
1},S,N)$.  We will often omit $S$ and $N$ in the arguments
henceforth to avoid clutter.  Since $R$ and $T$ satisfy the same
loop equations as $\Rt$ and $\Tt$  provided $g_1$ is chosen
appropriately, i.e. $g_1=g_1(g_{p\ge 2},S,N)\equiv \gt_1$, it
should be that
\begin{align}
 \Rt(z;g_{p\ge 2})=R(z;g_{p\ge 1}) \big|_{g_1=\gt_1},\qquad
 \Tt(z;g_{p\ge 2})=T(z;g_{p\ge 1}) \big|_{g_1=\gt_1}.
 \label{eq:R=Rt,T=Tt}
\end{align}
These satisfy the conditions (\ref{eq:contour_cond}) given that
$R$ and $T$ satisfy the conditions (\ref{eq:contour_cond}) without
tildes. Expanding these  in $z$, we find
\begin{align}
 \bracket{\Tr[\CW^2\Phit^p]}{}_{g_{p\ge 2}}^{\rm traceless}
 =
 \bracket{\Tr[\CW^2\Phi^p]}{}_{g_{p\ge 1}}^{\rm traceful} \big|_{g_1=\gt_1},
 \qquad
 \bracket{\Tr[\Phit^p]}{}_{g_{p\ge 2}}^{\rm traceless}
 =
 \bracket{\Tr[\Phi^p]}{}_{g_{p\ge 1}}^{\rm traceful} \big|_{g_1=\gt_1}.
 \label{eq:equivalence_of_trPhi}
\end{align}
In particular, setting $p=1$ in the second equation,
\begin{align}
 \bracket{\Tr[\Phi]}{}_{g_{p\ge 1}}^{\rm traceful} \big|_{g_1=\gt_1}
 =
 \left.\left[\frac{\partial}{\partial g_1}T(z;g_{p\ge 1})\right]\right|_{g_1=\gt_1}
 =
 0,
 \label{eq:tracelessness_Rt}
\end{align}
which can be used for determining $g_1$ in terms of all other parameters.%
\footnote{
One might have expected that $g_1$ can be determined by Eq.\
(\ref{eq:def_g1t}).
However, it is easy to show using the relation
(\ref{eq:equivalence_of_trPhi}) that the equation is just the equation
of motion of the traceful theory, which is identically satisfied for
any $g_1$:
$
 0
 \equiv \Bracket{\Tr[W'(\Phi)]}
 =\bracket{\Tr[\Wt'(\Phit)]}+N g_1.
$
}
We infer from Eq.\ (\ref{eq:R=Rt,T=Tt}) equality between the traceless
and traceful effective superpotentials:
\begin{align}
 \Wt_{\rm eff}(g_{p\ge 2},S,N)
 =
 W_{\rm eff}(g_{p\ge 1},S,N)|_{g_1=\gt_1(g_{p\ge 2},S,N)}.
\end{align}
As long as we impose the tracelessness condition
(\ref{eq:tracelessness_Rt}), this correctly reproduces the
relation (\ref{eq:equivalence_of_trPhi}).  Note that $\gt_1$
depends on $N$; this is the origin of the complicated $N$
dependence of $\Wt_{\rm eff}$ found in \cite{Kraus:2003jf}.

Because we know that the traceful theory can be solved by the associated
traceful matrix model, we can calculate the effective superpotential
using that matrix model.  Specifically, in the present case, it is given
in terms of the free energy of the traceful matrix model by
\begin{align}
 \Wt_{\rm eff}(g_{p\ge 2},S,N)
 =
 \left.\left[
 N\frac{\partial}{\partial S}\CF_{S^2}
 \right]\right|_{g_1=\gt_1}.
\end{align}
The function $\gt_1(g_2,g_3,\cdots,S,N)$ is determined by
\begin{align}
 \bracket{\Tr[\Phit]}
 =
 \left.\left[
 N\frac{\partial}{\partial S}\frac{\partial}{\partial g_1}
 \CF_{S^2}
 \right]\right|_{g_1=\gt_1}
 =
 0.
\end{align}

If we add fundamental fields, the shift constant $g_1$ is changed to
\begin{align}
 g_1 \equiv -\frac{1}{N}\Bracket{\Tr[\Wt'(\Phit)]}
 -\frac{1}{N}\Bracket{\Qt_\ft m_{\ft f}Q_f},
\end{align}
but everything else remains the same; we just have to work with
the traceful theory and the shifted tree level superpotential.
$g_1$ is determined by the tracelessness condition.

We only discussed the $SU(N)$ case in the above, but the
generalization to other tensors, i.e., $SO$ traceless symmetric
tensor and $Sp$ traceless antisymmetric tensor, is
straightforward.  We just shift the tree level superpotential as
(\ref{eq:shifted_supoerpot_traceless}), and work with the traceful
theory instead.

\subsection{Examples}

Here we explicitly demonstrate how the method outlined above works 
in the case of a cubic tree level superpotential, 
\begin{align}
 \Wt(\Phit)=\frac{m}{2}\Phit^2+\frac{g}{3}\Phit^3.
\end{align}
The associated traceful tree level superpotential is
\begin{align}
 W(\Phi)=\lambda \Phi+\frac{m}{2}\Phi^2+\frac{g}{3}\Phi^3
\end{align}
($g_1=\lambda,g_2=m,g_3=g$).

\subsubsection{SU(N) adjoint}

We first  consider $SU(N)$ with adjoint matter and no
fundamentals. 
In \cite{Kraus:2003jf} it was found by perturbative computation
to order $g^6$ that the corresponding $W_{{\rm eff}}$ vanishes due
to a cancellation among diagrams.  We will now prove that 
$W_{{\rm eff}} = 0$ to all orders in $g$.

The planar contributions to the free energy of the 
traceful matrix model can be
computed exactly by the standard method \cite{BrezinSV}:
\begin{align}
\CF_{S^2}
 =& S W_0 +\frac{1}{2} S^2  \ln\left( \frac{\tilde{m}}{ \sqrt{1+y}m}\right)
-\frac{2}{ 3} \frac{S^2 }{ y}
\left[1+\frac{3}{ 2} y
+\frac{1}{ 8} y^2-(1+y)^{3/2}\right]
\end{align}
with
\begin{align}
& \tilde{m}  = \sqrt{m^2-4 \lambda g} \nonumber \\
& W_0  = \frac{1}{2g} (\tilde{m}-m)\left(\lambda +\frac{1}{12g}
(\tilde{m}-m)(\tilde{m}+2m)\right) \nonumber \\
&\frac{y}{(1+y)^{3/2}}  = \frac{8 g^2 S}{m^3}.
\end{align}
We discarded some $g$ independent contributions.  The $W_0$ term 
arises from shifting $\Phi$ to eliminate the linear term in $W(\Phi)$. 
The superpotential is therefore
\begin{align}
W_{{\rm eff}}= N \frac{\partial \CF_{S^2}}{ \partial S}
=  N W_0  +  \frac{NS}{ 6 y}
 \left[-4-6y+6y \ln\left(\frac{\tilde{m}}{ m
\sqrt{1+y}}\right)+4(1+y)^{3/2}\right].
\label{Weff}
\end{align}
Imposing $\partial W_{{\rm eff}} / \partial \lambda =0$ leads to,
after some algebra,
\begin{align}
\lambda = - \frac{2 g S}{m}, \quad y = \frac{8 g^2 S}{m^3}.
\end{align}
Substituting back into (\ref{Weff}) and doing some more algebra, we find
\begin{align}
W_{{\rm eff}} = 0.
\end{align}
This vanishing of the perturbative contribution to the effective
superpotential is consistent with the gauge theory analysis of
\cite{Ferrari:2002jp}.   In fact, it is shown there that $W_{{\rm eff}} = 0$
for any tree level superpotential with only odd power interactions. 


\subsubsection{Sp(N) antisymmetric tensor}

Now consider $Sp(N)$ with an antisymmetric tensor and no
fundamentals. 
By diagram calculations or by computer, 
the planar and $RP^2$ contributions to the free energy of 
the traceful matrix model are 
\begin{align}
 \CF_{S^2}
 =&
 -\frac{\lambda^2S}{2 m}
 -
 \left(\frac{\lambda S^2}{2m^2}+\frac{\lambda^3 S}{3m^3}\right)g
 -
 \left(\frac{S^3}{6m^3}+\frac{\lambda^2 S^2}{m^4}+\frac{\lambda^4S}{2m^3}\right)
 g^2
 \nonumber\\
 &-
 \left(\frac{\lambda S^3}{m^5}+\frac{8\lambda^3 S^2}{3m^6}
   +\frac{\lambda^5 S}{m^7}\right)
 g^3
 -
  \left(
   \frac{S^4}{3m^6}
  +\frac{5 \lambda^2S^3}{m^7}
  +\frac{8 \lambda^4S^2}{m^8}
  +\frac{7\lambda^6S}{3m^9}
 \right)g^4
 \nonumber\\
 &
 -
  \left(
   \frac{4 \lambda S^4 }{m^8}
  +\frac{70 \lambda^3 S^3}{3 m^9}
  +\frac{128 \lambda^5 S^2 }{5 m^{10}}
  +\frac{6\lambda^7 S}{m^{11}}
 \right)
 g^5
 \nonumber\\
 &
 -
  \left(
   \frac{7S^5}{6m^9}
  + \frac{32\lambda^2 S^4}{m^{10}}
  + \frac{105 \lambda^4 S^3}{m^{11}}
  + \frac{256 \lambda^6 S^2}{3m^{12}}
  + \frac{33 \lambda^8 S}{2m^{13}}
 \right)
 g^6
 \nonumber\\
 &
 -
  \left(
   \frac{21 \lambda S^5}{m^{11}}
  + \frac{640 \lambda^3 S^4}{3m^{12}}
  + \frac{462 \lambda^5 S^3}{m^{13}}
  + \frac{2048 \lambda^7 S^2}{7m^{14}}
  + \frac{143 \lambda^9 S}{3m^{15}}
 \right)
 g^7
 \nonumber\\
 &
 -
  \left(
    \frac{16              S^6}{3 m^{12}}
   + \frac{231 \lambda^2   S^5}{ m^{13}}
   + \frac{1280 \lambda^4  S^4}{ m^{14}}
   + \frac{2002 \lambda^6  S^3}{ m^{15}}
   + \frac{1024 \lambda^8  S^2}{ m^{16}}
   + \frac{143  \lambda^{10}S^1}{ m^{17}}
 \right)
 g^8
 +\CO(g^9),
 \label{eq:corr_traceful_F_S2_Sp(N)_mm}
\end{align}
\begin{align}
 \CF_{RP^2}
 =&
 \frac{\lambda S}{2 m^2}g
 +
 \left(\frac{3\lambda S^2}{8m^3}+\frac{\lambda^2 S}{m^4}\right)g^2
 +
 \left(\frac{9\lambda S^2}{4m^5}+\frac{8\lambda^3 S}{3m^6}\right)
 g^3
 \nonumber\\
 &+
  \left(
   \frac{59 S^3}{48 m^6}
  +\frac{45\lambda^2 S^2}{4m^7}
  +\frac{8 \lambda^4 S}{m^8}
 \right)g^4
 +
  \left(
   \frac{59 \lambda S^3}{4m^8}
   +\frac{105 \lambda^3 S^2}{2m^9}
   +\frac{128 \lambda^5 S}{5m^{10}}
 \right)g^5
 \nonumber\\
 &+
  \left(
   \frac{197 S^4}{32m^9}
   +\frac{118\lambda^2 S^3}{m^{10}}
   +\frac{945 \lambda^4 S^2}{4m^{11}}
   +\frac{256 \lambda^6 S}{3m^{12}}
 \right)g^6
 \nonumber\\
 &+
  \left(
   \frac{1773 \lambda S^4}{16m^{11}}
   +\frac{2360 \lambda^3 S^3}{3 m^{12}}
   +\frac{2079 \lambda^5 S^2}{2m^{13}}
   +\frac{2048\lambda^7 S}{7m^{14}}
 \right)g^7
 \nonumber\\
 &+
  \left(
    \frac{4775           S^5}{128 m^{12}}
   +\frac{19503\lambda^2 S^4}{16  m^{13}}
   +\frac{4720 \lambda^4 S^3}{    m^{14}}
   +\frac{9009 \lambda^6 S^2}{2   m^{15}}
   +\frac{1024 \lambda^8 S  }{    m^{16}}
 \right)g^8
 +\CO(g^9),
 \label{eq:corr_traceful_F_RP2_Sp(N)_mm}
\end{align}
up to a $\lambda$ and $g$ independent part.
From the tracelessness (\ref{eq:tracelessness_Rt}), we find
\begin{align}
 \lambda
 =&
 \left(-1+\frac{2}{N}\right)\frac{S}{m}g
 +\left(-{3\over N}+{12\over N^2}\right)\frac{S^2}{m^4}g^3
 +\left(-{1\over N}-{24\over N^2}+{160\over N^3}\right)\frac{S^3}{m^7}g^5
 \nonumber\\
 &+
 \left(-{3\over 4 N}-{27\over N^2}-{192\over N^3}+{2688\over N^4}
  \right)\frac{S^4}{m^{10}}g^7
 +\CO(g^9)
 \equiv \lambdat.
\end{align}
Therefore, the effective superpotential is, up to an $\alpha$
independent additive part,
\begin{align}
 \Wt_{\rm eff}
 =&
 W_{\rm eff}|_{\lambda=\lambdat}
 =
 \left.\left[
   N\frac{\partial}{\partial S}\CF_{S^2} + 4\CF_{RP^2}
 \right]\right|_{\lambda=\lambdat}
 \nonumber\\
 =&
 \left(-1+\frac{4}{N}\right)S^2\alpha
 +\left(-\frac{1}{3}-\frac{8}{N}+\frac{160}{3N^2}\right)S^3\alpha^2
 +\left(-\frac{1}{3}-\frac{12}{N}-\frac{256}{3N^2}+\frac{3584}{3N^3}\right)S^4\alpha^3
 \nonumber\\
 &+\left(-\frac{1}{2}-\frac{24}{N}-\frac{352}{N^2}+\frac{33792}{N^4}\right)S^5\alpha^4
  +\cdots,
\end{align}
where $\alpha\equiv\frac{g^2}{2m^3}$.  This reproduces the result of
\cite{Kraus:2003jf} up to $\CO(\alpha^3)$ and extends it further to
$\CO(\alpha^4)$.

From these examples, the advantage of the present approach over the
traceless diagram approach of \cite{Kraus:2003jf} should be clear.  In
that approach, one has to evaluate contributing diagrams order by order
and evaluating the combinatorics gets very cumbersome.  On the other
hand, in this traceful approach, there is no issue of keeping and
dropping diagrams, and calculations can be done more systematically.
Therefore, being able to reduce the traceless problem to a traceful
problem is a great advantage.

\subsection{Traceless matrix model}

\begin{figure}[h]
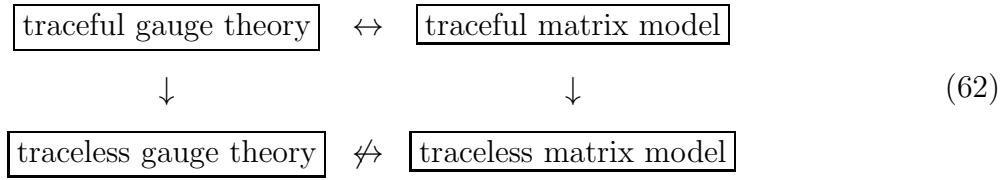

 \begin{align}
 \begin{array}{ccc}
  \fbox{traceful gauge theory} & \leftrightarrow & \fbox{traceful matrix model}
   \\[1ex]
  \downarrow && \downarrow \\[1ex]
  \fbox{traceless gauge theory} &\not\leftrightarrow & \fbox{traceless
  matrix model}
 \end{array}
 \end{align}
 \caption{Relation among traceful and traceless theories.}
 \label{fig:rel_traceless/traceful_gt_mm}
\end{figure}

We saw that the traceless gauge theory can be solved by the traceful
matrix model, not the traceless matrix model.  In the following, we
argue that the traceless matrix model is not useful in determining the
effective superpotential of the traceless gauge theory, $\Wt_{\rm eff}$.
The relation among traceless and traceful theories, as far as the
effective superpotential is concerned, is shown in Fig.\
\ref{fig:rel_traceless/traceful_gt_mm}.

%
%

The matrix model loop equation for traceless matter can be derived
almost in parallel to the traceless gauge theory loop equation derived
in the previous subsection.  Again, we replace the projector $P$ with
the appropriate traceless version $\Pt$.  For example, in the case of
$SU(N)$ adjoint without fundamentals, which was considered in the
previous section on the gauge theory side, the loop equation is
\begin{align}
 [W'\mRt_{S^2}]_- &= (\mRt_{S^2}{})^2.
 \label{eq:traceless_loop_eq_mm}
\end{align}
Here $W$ is the shifted superpotential defined in
(\ref{eq:shifted_supoerpot_traceless}), with $g_1$ defined in
(\ref{eq:def_g1t}) and the gauge theory expectation values replaced by
the matrix model expectation values.

Eq.\ (\ref{eq:traceless_loop_eq_mm}) is of the same form as the traceful
matrix model loop equation, and the first equation of the traceless gauge
theory loop equations (\ref{eq:traceless_loop_eq_gt}).  Finally, 
using the equivalence of the traceful gauge theory and matrix model, we
conclude that
\begin{align}
 \mRt(z;g_{p\ge 2})
 = \mR(z;g_{p\ge 1}) \big|_{g_1=\gt_1}
 = R(z;g_{p\ge 1}) \big|_{g_1=\gt_1}
 = \Rt(z;g_{p\ge 2}).
\end{align}
However, what we need to determine $\Wt_{\rm eff}$ is $\Tt$, which we
saw in the last subsection to be obtainable from the traceful theory as
\begin{align}
 \Tt(z;g_{p\ge 2},S,N)
 =
 T(z;g_{p\ge 2},S,N) \big|_{g_1=\gt_1(g_2,g_3,\cdots,S,N)}
 = \left.\left[
 N\frac{\partial}{\partial S}R(z;g_{p\ge 2})
 \right]\right|_{g_1=\gt_1(g_2,g_3,\cdots,S,N)}.
 \label{eq:Tt_ito_R}
\end{align} From the standpoint of the traceless matrix model, the only thing we
know is $\mRt=\Rt=R|_{g_1=\gt_1}$, and we have no information about the
$g_1$ dependence of $R$.  In the framework of the traceless matrix
model, there is no way of performing the derivative $\partial/\partial
S$ in (\ref{eq:Tt_ito_R}) before making the replacement $g_1=\gt_1$,
because $\gt_1$ depends on $S$ also.

Therefore, it is impossible to obtain the effective superpotential for
the traceless gauge theory directly,
just by using the data from the corresponding
traceless matrix model.  
We really need to invoke the traceful matrix model.


\section*{Acknowledgments}

%
This work was supported in part by NSF grant PHY-0099590.


\bigskip

\bigskip\bigskip\bigskip\noindent
{\LARGE \bf \appendixname}

\appendix

\section{Loop equations on the gauge theory side}
\label{app:loop_eq_gt}

In this appendix, we are going to calculate the gauge theory loop equations 
using the approach of \cite{Cachazo:2002ry} \cite{Seiberg:2002jq}. 
We start with 
generalized Konishi currents and corresponding 
transformations of the fields 
\ba
\label{eq: most generalized currents}
\begin{matrix}
J_f &\equiv& 
\Tr \Phi^\dagger e^{V_{\rm adj}} f (\WW_\alpha, \Phi) 
\hfill
&
\quad
\so 
\quad
&
\delta \Phi &=& f (\WW_\alpha, \Phi) 
\hfill
\cr
J_g &\equiv& 
Q^\dagger_f e^{V_{\rm fund}} g_{f f'} (\Phi) Q_{f'} 
&
\quad
\so 
\quad
&
\delta Q_{f} &=& g_{f f'} (\Phi) Q_{f'} 
\end{matrix}
\ea 
The explicitly written indices on the $Q_f$'s and $g_{f f'}$ are 
flavor indices, and gauge indices are suppressed.  
We find the generalized anomaly equations 
\ba
\label{eq: generalized anomaly}
\bar D^2 J_f
&=& 
\Tr f(\WW_\alpha , \Phi) W'(\Phi)
+
\tilde Q f(\WW_\alpha , \Phi) m'(\Phi) Q
+
\sum_{jklm} 
A_{jk,lm} 
{\partial f_{kj}
\over
\partial \Phi_{lm} }
\nonumber\\
\bar D^2 J_g
&=& 
2 
\tilde Q m(\Phi) g(\Phi) Q
+
\Tr A^{\rm fund} g(\Phi)
\ea
and $\bar D^2 J_f$ and $\bar D^2 J_g$ 
vanish in the chiral ring.

The field $\Phi$ being considered transforms by commutation 
under gauge transformations, 
so the elementary anomaly coefficient 
is the same as the one appearing in \cite{Cachazo:2002ry},  
\ba
\label{eq: elementary anomaly coefficient}
A_{jk,lm} 
&=& 
{1 \over 32 \pi^2} 
\left[
(\WW_\alpha \WW^\alpha)_{jm} \delta_{lk} 
+
(\WW_\alpha \WW^\alpha)_{lk} \delta_{jm} 
-
2 (\WW_\alpha)_{jm} (\WW^\alpha)_{lk} 
\right]
\hspace{-1.5ex}
\nonumber\\
&\equiv&
{1 \over 32 \pi^2} 
\left\{
\WW_\alpha , \left[ 
\WW^\alpha, e_{ml} 
\right]
\right\}_{jk}
\ea
where $e_{ml}$ is the basis matrix with the single non-zero entry
$(e_{ml})_{jk} = \delta_{mj} \delta_{lk}$. 
For fields transforming 
in the fundamental representation we should use 
\ba
\label{eq: elementary anomaly coefficient: fundamental}
A_{jk}^{\rm fund} 
&=& 
{1 \over 32 \pi^2} 
(\WW_\alpha \WW^\alpha)_{jk} 
\ea

There is one modification in the treatment of fundamental fields,
as compared to the $U(N)$ case studied in \cite{Seiberg:2002jq}.
Since the fundamental representation is real for $SO$ and
pseudo-real for $Sp$, the fields $Q$ and $\tilde Q$ are not
independent; instead, they are related by (\ref{eq:def_Qtilde}).
This results in the factor of 2 in the second equation in
(\ref{eq: generalized anomaly}), but otherwise the discussion
proceeds as in \cite{Seiberg:2002jq}. In the rest of the Appendix
we omit reference to fundamentals.

Next we consider the symmetries of $\Phi$. 
In equation (\ref{eq: most generalized currents}), 
$f = \delta \Phi$ 
must have the same symmetry properties as $\Phi$ itself. 
The tensor field will be taken either symmetric or antisymmetric.
We can discuss all four cases in a uniform fashion 
by using the notation 
\ba
\Phi^T = 
\left\{
\begin{matrix}
\sigma \Phi
\hfill
&
\quad
\mbox{for groups $SO(N)$,}
\cr
\sigma J \Phi J^{-1}
&\quad
\mbox{for groups $Sp(N)$,}
\end{matrix}
\right.
\ea
and $\sigma = \pm 1$. 
The gauge field satisfies 
$\WW_\alpha{}^T = - \WW_\alpha$ for $SO$ groups, and 
$\WW_\alpha{}^T = - J \WW_\alpha J^{-1}$ for $Sp$ groups. 
%
%
As discussed in Subsection \ref{section: loop equations}, $\Phi$
has the property \ba \label{eq: projector: general} \Phi = P \Phi
,\quad\mbox{or explicitly}\quad \Phi_{ab} = P_{ab,ij} \Phi_{ij}
\ea with the projectors defined in (\ref{eq:def_projector}). To
ensure that $f$ has the same symmetry as $\Phi$, we should replace
$f \to P f$ in (\ref{eq: generalized anomaly}). Specifically, we
will take $\delta \Phi$ of the form \ba \label{eq: delta Phi:so}
f^{SO} &=& P^{SO} {B \over z - \Phi} = \left( {B \over z - \Phi}
\right) + \sigma \left( {B \over z - \Phi} \right)^T = \left( {B
\over z - \Phi} \right) + \sigma \left( {B^T \over z - \sigma
\Phi} \right) ,
\\
\label{eq: delta Phi:sp}
f^{Sp} &=& 
P^{Sp} {B \over z - \Phi} =
\left( {B \over z - \Phi} \right) 
+ \sigma J \left( {B \over z - \Phi} \right)^T J
= 
\left( {B \over z - \Phi} \right) 
+ \sigma \left( {J B^T J \over z -\sigma \Phi} \right)
,
\ea
with $B = 1$ or $B = \WW^2 \equiv \WW_\beta \WW^\beta$. 
Using the symmetry of the gauge field
and the chiral ring relations, 
both (\ref{eq: delta Phi:so}) and (\ref{eq: delta Phi:sp}) 
reduce to 
\ba
\label{eq: delta Phi: in CR only}
f &=& 
{B \over z - \Phi}  
+ \sigma 
{B \over z - \sigma \Phi} 
.
\ea
%
%
Also, to take derivatives with respect to matrix elements%
\footnote{
    In the case of $U(N)$ of \cite{Cachazo:2002ry},
    one had $P_{lm,ab} = (e_{lm})_{ab}$
    which satisfies $(e_{lm})_{ab} B_{ab} = B_{lm}$, for any matrix $B$.
    }
correctly we should set 
\ba
\label{eq: change of matrix elements}
\partial_{lm} \Phi_{ab} 
= 
P_{lm,ab}
\ea
Then the tensor field anomaly term becomes 
\ba
\label{eq: anomaly: general}
A_{jk,lm} 
\; 
\partial_{lm} f_{kj}
&=& {1 \over 32 \pi^2} \left[ (\WW^2)_{jm} \delta_{lk} +
\delta_{jm} (\WW^2)_{lk} - 2 (\WW_\alpha)_{jm} (\WW^\alpha)_{lk}
\right] \nonumber\\&&\hspace{1.4em}\times \left[ \left( {B \over z
- \Phi} \right)_{ra} \left( {1 \over z - \Phi} \right)_{bs}
P_{kj,rs} P_{lm,ab} \right]. \ea After using the projectors
(\ref{eq:def_projector}), the identity $\Tr \WW_\alpha \Phi^k =
0$, the symmetry properties of $\Phi$ and $\WW_\alpha$, and the
chiral ring relations, we find \ba \label{eq: anomaly: simplified}
A_{jk,lm} \;
\partial_{lm} f_{kj}
&=& {1 \over 32 \pi^2} \left[ \left( \Tr {\WW^2 \over z - \Phi}
\right) \left( \Tr {B \over z - \Phi} \right) + \left( \Tr {1
\over z - \Phi} \right) \left( \Tr {\WW^2 B \over z - \Phi}
\right) \right.\nonumber\\&&\left.\hspace{3em} + 4 k \sigma \; \Tr
{\WW^2 B \over (z - \Phi) (z - \sigma \Phi)} \right] \ea The only
difference in (\ref{eq: anomaly: simplified}) between the two
types of gauge groups is that the sign in front of the single
trace term is $k = +1$ for $SO$, and $k = -1$ for $Sp$.
Taking $B = 1$ and $B = \WW^2$ in (\ref{eq: anomaly: simplified}) we find 
\ba
\label{eq: anomaly: simplified: B=1}
0 &=& 
\Tr { W'(\Phi) \over z - \Phi}
+ \sigma 
\Tr { W'(\Phi) \over z - \sigma \Phi}
\nonumber\\&&
+
{2 \over 32 \pi^2} 
\left[
\left( \Tr {\WW^2 \over z - \Phi} \right) 
\left( \Tr {1 \over z - \Phi} \right) 
+ 2 k \sigma 
\left( \Tr {\WW^2 \over (z - \Phi)(z - \sigma \Phi)} \right) 
\right]
\\
\label{eq: anomaly: simplified: B=W-squared}
0
&=& 
\Tr { \WW^2 W'(\Phi) \over z - \Phi}
+ \sigma 
\Tr { \WW^2 W'(\Phi) \over z + \Phi}
+ {1 \over 32 \pi^2} \left[ \left( \Tr {\WW^2 \over z - \Phi}
\right) \left( \Tr {\WW^2 \over z - \Phi} \right) \right] \ea Now
recall that $W(\Phi)^T = W(\Phi)$ for $SO(N)$, and $W(\Phi)^T = J
W(\Phi) J^{-1}$ for $Sp(N)$ since it only appears inside a trace;
so \ba \Tr { W'(\Phi) \over z -\sigma \Phi} &=& \sigma \Tr {
W'(\Phi) \over z - \Phi} ,\quad \Tr { \WW^2 W'(\Phi) \over z
-\sigma \Phi} = \sigma\Tr { \WW^2 W'(\Phi) \over z - \Phi} . \ea
The single trace terms have to be treated separately: when $\sigma
= -1$, \ba \Tr { \WW^2 \over z^2 - \Phi^2} &=& {1\over 2 z} \; \Tr
\left[ \WW^2 \left( { 1 \over z - \Phi} + { 1 \over z + \Phi}
\right) \right] = {1\over z} \; \Tr { \WW^2 \over z - \Phi} \ea
while for $\sigma = +1$, we should use \ba \Tr { \WW^2 \over (z -
\Phi)^2} &=& - {d\over dz} \; \Tr { \WW^2 \over z - \Phi} . \ea

Putting everything together, we find the loop equations written in
equation (\ref{eq:loop_eq_gt}).

\section{Loop equations on the matrix model side}
\label{app:loop_eq_mm}

Here we derive the matrix model loop equations for $SO/Sp$
following Seiberg \cite{Seiberg:2002jq}, who discussed the $U(N)$
case.  Start with the matrix model partition function
\begin{eqnarray}
\label{eq: mm partition function}
Z = \int d \mPhi d \mQ\,
\exp 
\left\{
- {1 \over \mg} \left[
\Tr[W(\mPhi)]
+ 
\mQt_{\tilde f} m_{\tilde f f}(\mPhi) \mQ_f\right]
\right\}.
\end{eqnarray}
Because the fundamental matter is real for $SO(N)$ and pseudo-real
for $Sp(N)$, there is no integration over $\mQt$.  It is not an
independent variable, but related to $\mQ$ by Eq.\
(\ref{eq:def_Qtilde}).  We will write the symmetry properties of
the the tensor field $\mPhi$ as
\begin{eqnarray}
\mPhi^T = 
 \begin{cases}
  \sigma \mPhi & SO(N), \\
  \sigma J\mPhi J^{-1} & Sp(N). 
 \end{cases}
\end{eqnarray}
where $\sigma = \pm 1$.  The matrix $m(\mPhi)$ has symmetry properties as
given in Eq.\ (\ref{eq:sym_prop_m}).

Now we perform two independent transformations 
\begin{eqnarray}
\label{eq: transformations: mm}
\delta \mPhi = B P {1 \over z - \mPhi}
,\quad
\delta \mQ_f = \lambda_{f f'} {1 \over z - \mPhi} \mQ_{f'}
\end{eqnarray}
where $B$ (number) and $\lambda$ (matrix) are independent and
infinitesimal.  
To make sure that $\delta \mPhi$ has the same 
symmetry properties as $\mPhi$ itself, we have introduced 
the appropriate projector $P$ in (\ref{eq: transformations: mm}), 
see Eq.\ (\ref{eq:def_projector}).  
The measure in
(\ref{eq: mm partition function}) changes as
\begin{eqnarray}
d \mPhi &\to& d\mPhi\, J_\mPhi = d\mPhi\, (1 + \Delta_\mPhi),
\nonumber\\
d \mQ &\to& d\mQ\, J_\mQ = d\mQ\, (1 + \Delta_\mQ)
\end{eqnarray}
to first order in $B$ and $\lambda$, 
where the corresponding changes in the Jacobians are 
\begin{eqnarray}
\Delta_\mPhi &=& 
B 
P_{ij,ab} 
\left( {1 \over z - \mPhi} \right)_{ia} 
\left( {1 \over z - \mPhi} \right)_{bj}
\nonumber\\
\Delta_\mQ &=& 
\left( \Tr {1 \over z - \mPhi} \right)
\left( \tr \lambda \right),
\end{eqnarray}
where $\tr$ is a trace over the flavor indices.
The classical pieces change by 
\begin{eqnarray}
\delta \, \Tr[W(\mPhi) ]
= B \, \Tr\left[W'(\mPhi) P {1 \over z - \mPhi}\right]
= B \, \Tr {W'(\mPhi) \over z - \mPhi}
.
\end{eqnarray}
One can show the second equality using symmetry properties of 
$W$: since it only enters $Z$ in the form of the trace, we should 
take $W(\mPhi^T) = W(\mPhi)$ for $SO$ and 
$W(\mPhi^T) = J W(\mPhi) J^{-1}$ for $Sp$. 
Similarly, 
\begin{eqnarray}
\delta ( \mQt m \mQ )
&=& 
\mQt \left( {\lambda^T m \over z - \sigma \mPhi} 
+ {m \lambda \over z - \mPhi} \right)\mQ 
+ B \mQt m' \left( P {1 \over z - \mPhi} \right) \mQ 
\nonumber\\
&=&
2 \mQt {m \lambda \over z - \mPhi} \mQ 
+ B \mQt {m' \over z - \mPhi} \mQ 
,
\end{eqnarray}
where we used a similar symmetry property of the matrix $m$. 
Finally, with the explicit form of the projectors 
(\ref{eq:def_projector}) we find that in all four cases the 
statement $\delta Z = 0$ gives two independent loop equations 
(one for $B$, and one for $\lambda$): 
\begin{align}
\label{eq: exact mm loop equations}
{1\over 2}
\left\langle
\left(\mg
\Tr {1 \over z - \mPhi}
\right)^2
\right\rangle
\pm
{\sigma\over 2} {\mg}
\left\langle
{\mg} \Tr {1 \over (z - \mPhi)(z - \sigma \mPhi)}
\right\rangle
&=
\left\langle
{\mg} \Tr {W'(\mPhi) \over z - \mPhi}
\right\rangle
+ {\mg}
\left\langle
\mQt {m'(\mPhi) \over z - \mPhi} \mQ
\right\rangle,
\nonumber\\
\left\langle
{\mg} \Tr {1 \over z - \mPhi}
\right\rangle
\delta_{ff'}
&=
2
\left\langle
\mQt_{\ft} {m_{\ft f}(\mPhi)  \over z - \mPhi} \mQ_{f'}
\right\rangle,
\end{align}
for $SO$ and 
$Sp$, respectively.
This is Eq.\ (\ref{eq:exact_mm_loop_eqs}) 
quoted in Section \ref{section: mm loop equations}.

As it is written, equation (\ref{eq: exact mm loop equations}) 
includes all orders in $\mg$. The anomaly term in the first 
equation (\ref{eq: exact mm loop equations}) factorizes as 
\ba
\left\langle
\left(\mg
\Tr {1 \over z - \mPhi}
\right)^2
\right\rangle
= 
\left\langle
\mg
\Tr {1 \over z - \mPhi}
\right\rangle^2
\times \left[ 1 + \CO(\mg^2) \right]
\ea
as can be seen from a diagram expansion. 
With this and the definition of matrix model resolvents 
(\ref{eq:def_mm_resolvents}) and (\ref{eq:mm_resolvents_expansion}), 
we obtain the loop equations (\ref{eq:loop_eq_mm}).


\end{document}